# Assessment and Performance of Flexible Quench Antenna Array Diagnostics for Superconducting Magnets

Stoyan Stoynev and Joe DiMarco

*Abstract*— FNAL has been developing multiple versions of flexible quench antennas (flex-QA), including some specially optimized for high sensitivity and/or high resolution, to characterize quench events and transients during current ramping in superconducting magnets. A fundamental feature in our use of these is the creation of grid-like structures of sensitive elements to cover coil surfaces, with the aim of getting precise localization of magnetic flux-change events. The flex-QA are coupled with fast data-acquisition, allowing comprehensive analysis of signals at the relevant fine time scales. In addition to arrays of various flex-QA types being used during cryogenic testing of superconducting magnets, we also are utilizing a newly developed room temperature test stand to better understand QA response characteristics. The data from actual superconducting magnet tests, "warm" test stand measurements, and simulation data on the same QA designs allows us to draw conclusions on operational feasibility and plan better for improvements of our sensors. In this paper we present data from the multiple tests performed and analysis results. Flex-QA designs are compared, and their features, options, and optimization discussed.

*Index Terms*—Antenna arrays, electromagnetic measurements, electromagnetic transients, superconducting magnets.

## I. Introduction

Diagnostic tools are crucial in understanding performance of superconducting accelerator magnets for which the designs and technology continually develop. Instrumentation is needed which has both high spatial and temporal resolution, as well as a variety of sensor types in order to fully understand and describe the complex phenomena driving magnet performance - particularly affected by quenching and its dependencies. FNAL has developed new type of arrays based on flexible printed-circuit board technology and quench antenna sensing, flex-QA [1], to help address the diagnostic needs here. Earlier studies with individual small QA [1], [2] indicated significant potential of this approach and mixed-type array configurations for long magnets [3] also proved the concept viable. Flex-QAs from one of the newly designed arrays were embedded in a mirror magnet [4], following the curved aperture and facing the coil at ~ 5 mm distance. The same design was also tested later at a specially developed room temperature test stand [1], where tests were subsequently also completed on several other flex-QA designs.

In this paper, we describe the work on QA data acquired from a superconducting magnet, along with the resulting analysis and insights. We explore bench-testing techniques relatable to the experiment and compare to simulations. We additionally examine features of the different flex-QA arrays, obtained on the warm test stand, to guide us to improved designs. Main points are discussed, and conclusions are drawn.

## II. Superconducting Magnet Test with Flex-QA Arrays

### A. Flex-QA positioning and DAQ

The QA design with which we chose to instrument a "mirror" magnet is shown schematically in Fig. 1 and pictured in ([1], Fig. 5). It features 20 independent diagonal channels on a rectangular panel with sensitive area 80 mm x 460 mm. Two lengths of those cover the inner layer coil area, given the coil active area was ~ 80 cm long and had 60 mm aperture. Two panels were connected in series (20 channels total) and two others placed on top of them individually (40 channels together) to form a grid of 60 channels as presented in Fig. 1. Fig. 2 shows an actual picture of the flex-QAs with the 3D printed support seen under the panels. Features on the coil, QA arrangements and magnet parts (laminations) were all used together to relate geometrically the coil-to-QA coordinates with the goal of matching positions to better than 1 mm. The many twisted-pair wires coming out of the partial-bore of the magnet were carefully routed and labeled.

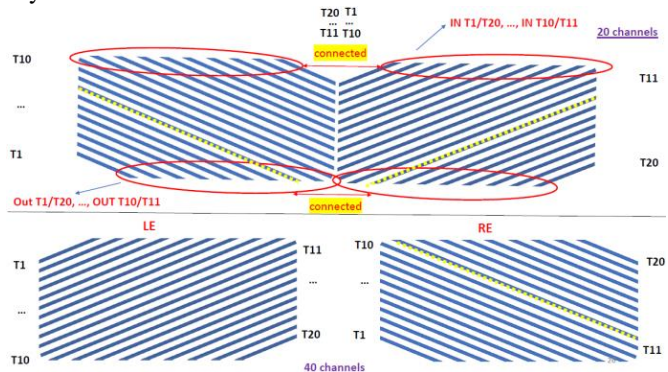

Fig. 1. Flex-QA channel arrangement. The top two shown have their QA sensors connected in series and the bottom ones (LE, RE) remain independent. Dotted lines show an example channel on each of the top and bottom QA.

This work was supported by Fermi Research Alliance, LLC, under Contract DE-AC02-07CH11359 with the U.S. Department of Energy, Office of Science, Office of High Energy Physics.(*Corresponding author: Stoyan Stoynev.*)

The authors are with Fermilab National Accelerator Laboratory, Batavia, IL 60510 USA (e-mail: stoyan@fnal.gov).

Color versions of one or more of the figures in this paper are available online at http://ieeexplore.ieee.org.

Digital Object Identifier will be inserted here upon acceptance.

The 60 QA signal channels along with the quench detection signal were read by a DAQ system based on NI-6143 modules (8 channels, 16-bit, 250 kHz). We took data at 100 kHz per channel rate. Data recording included the whole magnet current ramps, facilitated by LabView software.

### B. Magnet quenching and flex-QA data analysis

The magnet under consideration in this paper was part of other developments [4] and its test program started at 4.5 K. Most magnets quench predominantly in their coil inner layer(s) where the field is higher, and the first quench in our test was in the inner layer. This layer was instrumented with voltage taps and based on those we knew the likely quench location was somewhere around the second spacer/end-part, within ~40 cm cable length. We also had acoustic sensors installed on both magnet ends and, given that the first quench yielded fairly large mechanical signals, we triangulated the likely source to within 10 cm range in longitudinal direction.

With typical noise level at below 1 mV, we identified many QA signals above the noise at a distinctive time which we identified as quench time. Flex-QA quench time was typically few ms before visible voltage rise from the quenching voltage tap segment, as observed in previous work with flex-QA as well [1]. While QA signals show complex behavior over time, the first reaction to quench is a sharp peak(s) developing within 200 micro-seconds. Within our time resolution of 10 micro-seconds all channels featuring this peak reacted simultaneously. The magnitude of the peaks however differed, and we used it to associate highest level channels to geometrical origin of current redistribution. The highest peaks in quench one, 16 mV and 14 mV, were seen in channels shown on Figures 1 (dotted), 3 (highlighted) and 4 with 12 other channels reaching peak magnitudes of 2-8 mV. The precise quench location from QA was consistent with the rough voltage tap based constraints around the pole turn and the axial constraints from the acoustic signals. Thus, multiple diagnostics tools can be utilized for building a coherent and richer picture although VTs and acoustics were not quite sensitive enough to provide good location in this case.

Most of the quenches in this magnet test occurred in the outer layer, there were tens of those quenches. The flex-QA array was sensitive enough to register all of them – inner layer quenches went to an immediate-signal/noise ratio of over 35 while outer layer ones were at a ratio of up to 10. Fig 4. shows a zoomed in version of typical quench patterns. We identified four distinctive patterns, including time dependencies, and the one on the right was associated with almost all outer layer quenches. The channels with highest amplitudes at the time of quench (within 200 micro-seconds) point to a spot in the non-lead end of the outer layer, approximately consistent with coil damage regions as discussed in [4]. Fig. 5 is another geometrical representation of flex-QA array signals incorporating the need for correlated processing. It shows sensitivity at particular time slice, with sensitivity for cross-channels (1,2) defined as $s = \sqrt{M_1 \cdot M_2}$, where M is signal normalized to std. dev. computed "pre-quench" ([-0.2, -0.1] s) and based on running average over 1 ms. Noise filtering could improve sensitivity further but is not applied here. Those and other representations revealed a "weak" spot in the non-lead end outer layer associable to quenches. This area consistently sees current redistribution in adjacent QA channels up to the quench detection time. The same area was active between quench and quench detection times even in the first quench which was in the inner layer. Thus, evidence suggests the main damage to the coil outer layer was done at or before the first spontaneous quenching occurred.

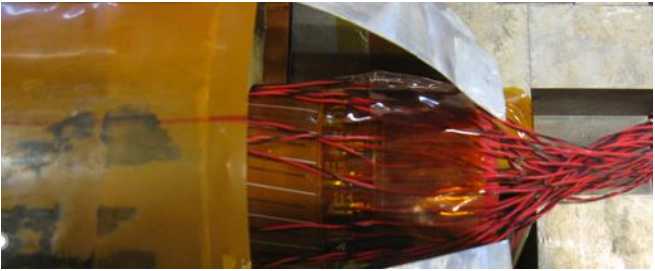

Fig. 2. Flex-QA arrays are placed on top of 3D-printed support seen under the bundle of wires coming from the QA. The coil is placed on top of them; Kapton insulation sheet and protection heaters complete the view. All components are placed on "mirror" iron blocks which are part of the "mirror" magnet.

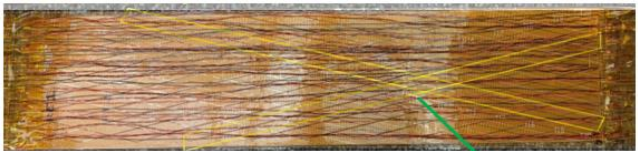
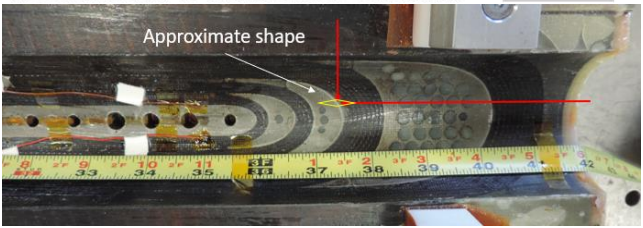

Fig. 3. Top: Superimposed images of QA arrays on top of each other with the two highest amplitude channels responding to the first quench highlighted. Bottom: Located position on the inner layer (ID) surface of the coil (note: the coil is seen from "below" and QA images are taken from "above").

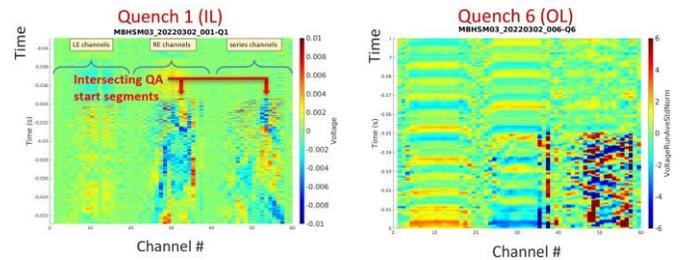

Fig. 4. QA array activity around quench time in first (inner layer /IL/, left) and the sixth (outer layer /OL/, right) magnet quenches. Noise patterns at 60 Hz are clearly visible and phase-shifted in the boards covering the two halves of the coils and are largely missing in the QA array consisting of two connected boards over the whole coil (channels 41-60). Absolute and normalized signals are shown for quench one and six, respectively; no filtering is applied.

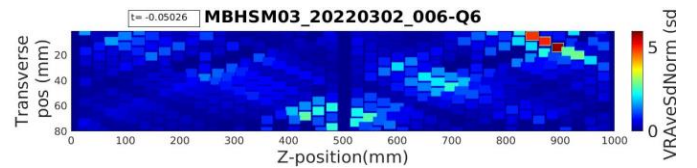

Fig. 5. Most outer layer quenches show this pattern at the first moments after quench start. The active area (reddish color) corresponds to locations between the end part and the midplane in the outer coil layer. Dimensions on the figure are approximate and used for visualization on the coil (which is in fact curved).

### III. FLEX-QA ARRAYS AT THE TEST-BENCH

#### A. Warm test stand characteristics and test goals

The QA warm test stand (WTS), [1] and Fig. 6 (left), allows to position and control a current source (loop) close to a QA board(s) under testing. After experimenting, we decided to adopt a loop of 2 cm diameter oriented in "Y-Z" plane (Y is along sensor length) emulating the cable plane at the straight section in a magnet during current redistribution and quench propagation. The current loop is fixed with the help of a shaper attached to the positioning system. The lowest point of it is maintained at ~ 1 mm from the QA surface during measurements, to allow for proper signal-to-noise ratio, Fig. 6 (right).

Flex-QA arrays were placed in horizontal positions aligned and fixed in the "X-Y" plane. To minimize electrical noise, we had to cover the surface under the QA board with aluminum foil and grounded it which eliminated any modulation in most cases. The signal noise level was as low as 0.2 mV although few individual channels reached ~1 mV modulated (i.e. still reducible) noise. The boards were placed parallel to the motion system coordinate frame within a very good degree, given the length of boards was ~ 0.5 m. However, the flatness of boards within parts of mm cannot be guaranteed. As the current source was 1 mm away, some "waviness", largely due to QA manufacturing and material features, may be expected.

A generator-controlled power supply was used to provide current of ±12 A in the form of square waves with 100 Hz frequency. Data was recorded with the same DAQ system used for the cold magnet test and same sampling rate of 100 kHz.

WTS measurements can serve multiple purposes. Response of QA channels are recorded and effectively used for relative calibration. Crosstalk effects are checked and investigated. Abnormal behavior in particular channels/board can also be pin-pointed. Current redistribution models or orientation dependencies, in the form of current loop(s) characteristics, can be studied. Most importantly, sensitivity within the QA arrays can be probed which affects design choices or applicability to cases. Finally, WTS can be utilized to emulate observed responses in real magnets, given a proper current shape is provided; and, ultimately, help to explain or predict real responses. Elaborated QA sets could be assessed there before committing to cold-test application while simulations could be benchmarked in advance of cold testing.

#### B. Test bench data

The flex-QA arrays we developed had two separate main features – inclined with respect to border lines sensors, a.k.a. "diagonal" design, and "triangle-pair" bucked design forming a rectangular sensor. We fabricated both, keeping the sensor width the same, 8 mm, and overall length and width of the different boards also the same ([1], Fig 5). The "diagonal" design, used in magnet testing, featured 20 sensors of variable length per board. The "triangle-pair" design consisted of 10 equivalent sensors, each having six identical sections of bucked "triangle-pairs", all sections connected in series.

Fig. 6 (right) shows QA response to a single current "jump"; the decay constant $\tau$, primarily driven by the system and not the QA channels, is ~ 0.2 ms. Thus, the half-period of signal modulation corresponds to 25 $\tau$.

The "diagonal" sensors have a clear symmetry defined by their mid-line and Fig. 7 confirms a dead-zone near the line of symmetry. By design this symmetry line is not along the current or quench propagation and is thus a limited "blind-spot" in practice. The response at channel edges is very strong which allows relatively strong readout of multiple channels at a given time/position, Fig 7 (bottom), when the source scans the board width, Fig. 7 (top). As long as noise or systematic effects are negligible, having multiple sensors providing scaled response is of utmost importance to pin-point current variation sources. Alternatively, these give a strong clue about acceptable sensor sizes, assuming emulation of current source relative location to QA sensor is satisfactory.

The "diagonal" QA array showed multiple occasions of reproducible "cross-talk": the noise floor (average read-out value without signal) is affected by signal in neighboring channel(s). We do not know yet the source of this systematic problem and it was not the case for all channels. We did not observe similar behavior before, in short or long single or multiple adjacent short channels. Crosstalk is a serious issue if one wants to use a multi-channel approach for providing better spatial resolution. It is instructive to note that the "triangle-pair" design showed no such problem but manufacturers, along with other parameters, were different between QA board designs.

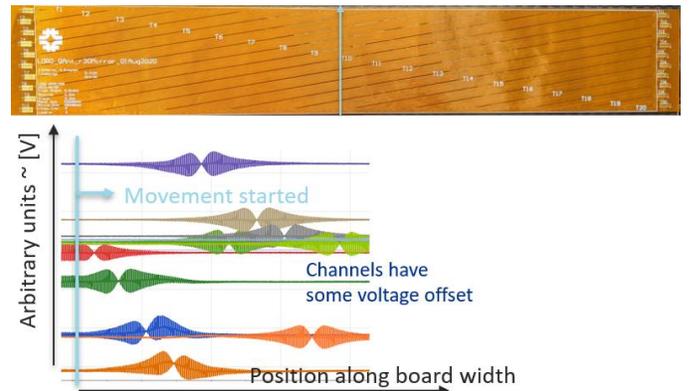

Fig. 7. Top: A scan path (blue arrow) over a "diagonal" design board. Dark lines are the middles of the sensors and pale bright lines in between indicate separation of channels. Channels are symmetric along the dark lines. Bottom: Results from the scan at constant speed. Inflection points (no signal) indicate crossing the symmetry lines (middles of channels) where signal vanishes as magnetic flux through the sensor on the left of the current loop compensates the flux on the right. Distance between those inflection points is nominally 8 mm / cos (0.165 rad). Non-symmetries with respect to the offset voltages indicate possible cross-talk. Voltage offsets are not channel ordered.

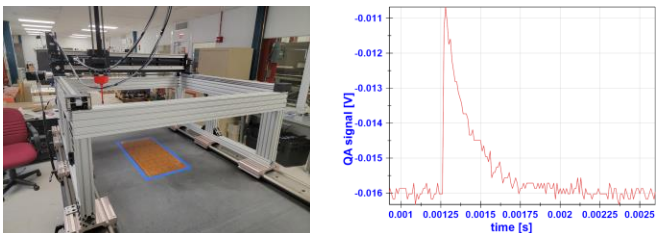

Fig. 6. Left: WTS overview with a flex-QA array ready for testing. Right: A QA channel response to current "jump" (-12 A to +12 A) – time constant of 0.2 ms is measured. Noise level (RMS) is typical; voltage offset is channel dependent and signal strength is determined with respect to it.

Fig. 8 shows scan directions across channels (comprised of multiple "sections") of the "triangle pair" design (top) and resulting signal (bottom). The response to signal above the middle part of a section (see Fig. 9 (left)) is particularly insensitive close to channel edges where an "island" is present at each edge; the adjacent channels are insensitive for the same reason. By design, current and quench propagation is not parallel to the "island" lines and thus dead-zones have limited practical length. Although not shown, inflection points (low signal) are also present between sections of a single channel. Sectioning cannot be avoided for long and narrow channels without introducing too long dead-zones in real applications. The advantages of the "triangle-pair" design relate to robustness against noise. A main disadvantage is that having less sensitivity along edges of channels is apparently worse suited for multi-channel use.

### C. Simulations

We explore simple simulation [3] where a current doublet with a length 1/5 of a channel section of the "triangle pair" design acts on the simulated version of the sensor. On Fig. 9 the pair is emphasized on a picture (left) and results from simulations shown along with the shapes from WTS data (right). In the simulation the doublet, approximating a current redistribution source in a magnet and resembling the WTS probe (i.e., 2 cm diameter over a 8-cm-long section), is moved along the width of the sensor at five adjacent positions along its length and fixed distance from the sensor surface. The simulation describes well how the signal changes from one side to the other over the sensor width/length and is consistent with data (in data we change signal polarity while moving the source; in the analytic simulation we calculate the absolute flux/voltage). It is apparent that behavior is reproducible and understandable to a good degree even though some parameters (e.g. round probe vs a straight doublet) deviate. It is a good start to tune simulations with different simple sources on various designs and then move to more elaborate source shapes which may describe the 3D-shapes of real magnet current redistribution better.

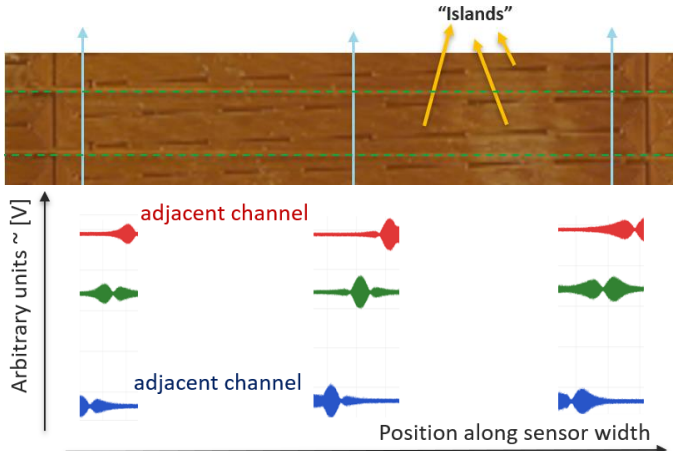

Fig. 8. Top: Scan paths (blue arrows left, middle, right) across the board (crossing over several channels). "Islands" (along lines) are the equivalents of the symmetry lines in the "diagonal" design – they correspond to lowest signal within a channel. Bottom: Results from the scans. Signal depends on source positions along both channel length and width.

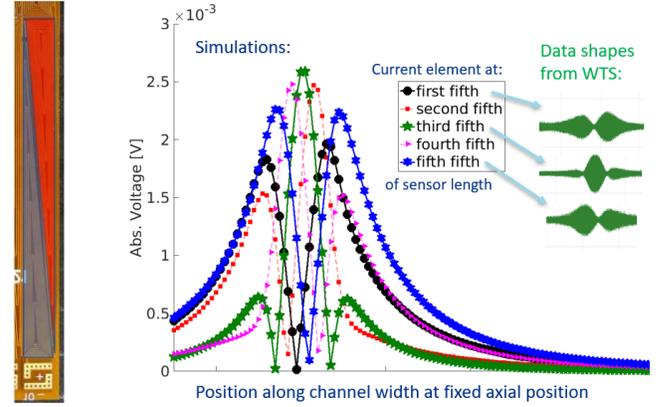

Fig. 9. Left: A triangle-pair section design portion of a sensor. Right: Simulated response of the circuit to current doublet at various positions (fixed axial, scan along the channel width). WTS data at relevant positions is compared to simulations. Only the main sensor geometry is embedded in simulations.

Simulations will have to be also employed to resolve convoluted signals in data. The initial spike we associate with quench start has the characteristic decay of our DAQ system (~0.2 s decay constant) but it is followed by other spikes and more gradual signal changes, many of them higher than the initial spike amplitude. At conceptual level this can be interpreted as consecutive fast current redistribution steps (each within tens to hundreds of micro-seconds), possibly from individual strands being shut off and on during current sharing and quenching [5]. Refined simulation studies are needed to describe the actual process with a good level of confidence.

## Conclusion

We demonstrate how a flex-QA arrays can be used to pinpoint quench locations with remarkable precision. They are very reliable, and signals are well above noise. We identified all quench starts in a superconducting magnet test well before quench detection and before any visible voltage rise in the conductor - this includes tens of quenches in the outer layer with the whole thickness of the inner coil layer between current redistribution location and the QA sensors. Our "warm" test stand studies indicated some weaknesses in the QA designs and possibly unavoidable imperfections but also showed that multi-channel sensitivity to quenches can serve as an additional tool to improve precision. Time development is yet to be fully explored along with more detailed studies on flex-QA designs at the warm test stand and in actual cryogenic magnet testing.

## Acknowledgments

We thank T. Cummings and N. Moibenko for QA design work, V. Nikolic for mechanical engineering leadership, O. Kiemschies and S. Krave for comprehensive DAQ support, Kelsey Scheidt for helping with WTS measurements.